\documentclass[showpacs,showkeys,preprintnumbers,amsmath,amssymb]{revtex4}
\usepackage{graphicx}
\begin{document}
\title {Adiabatic  Heavy  Ion Fusion Potentials for Fusion at Deep
Sub-barrier Energies}
\author{S. V. S. Sastry, S.Kailas, A. K. Mohanty and A. Saxena}
\affiliation{ Nuclear Physics Division, Bhabha Atomic Research Centre,
Trombay, Mumbai 400 085, India}
\begin{abstract}{
It  is shown that the energy (E) and angular momentum (L) dependent barrier
penetration model ( {\small{ELDBPM}} ) provides a consistent description of
the unusual behaviour of  fusion  cross  sections  at  extreme  sub-barrier
energies  reported recently. The adiabatic limit of fusion barriers derived
from experimental data are consistent with the  adiabatic  fusion  barriers
given    by    modified   Wilzynska-Wilzynski   prescription.   Using   the
{\small{ELDBPM}} , the fusion barrier systematics has been obtained  for  a
wide  range  of heavy ion systems. For some systems, the barrier parameters
R$_b$ and $\hbar\omega$ are anomalously smaller  than  the  predictions  of
known empirical prescriptions.
}\end{abstract}
\date  {\today}
\maketitle

\noindent
Heavy  ions  fusion  has been under intense investigation over the last two
decades. Heavy ions fusion populates compound  system  at  high  excitation
energies  and  angular  momenta,  providing  a tool to study the nucleus at
extreme conditions. It also opens up a new channel to generate  nuclei  far
away  from  stability line, populate the super heavy nuclei near the island
of stability. Further, the fusion of heavy  ions  at  sub  Coulomb  barrier
energies  exhibited anomalously large enhancements over the model estimates
based  on  tunneling  of  the  entrance  channel  Coulomb   barrier.   This
enhancement has been understood to be arising due to the strong coupling of
structure  of  the  fusing  nuclei and the reaction dynamics \cite{cc}. The
effects couplings of nuclear structure on fusion has been  clearly  brought
out  in  the  experimental  studies  of  the  fusion  barrier distributions
\cite{mdasgrev} and see the references in \cite{mdasgrev}. These  dynamical
aspects  of  heavy  ion reactions are successfully explained by the coupled
reaction channels ( {\small{CRC}} ) formalism, by a  simultaneous  solution
to  all  the reaction channels as in \cite{ijt1}. Nuclear fusion at extreme
sub-barrier energies has been  revisited  recently  \cite{ninisub,hagino1}.
Jiang  {\it  et.  al.} \cite{ninisub} have reported unexpected behaviour of
fusion cross sections at  extreme  sub-barrier  energies.  The  Wong  model
results  fail  to  reproduce  the  data  at  these  low energies. They have
observed that the logarithmic slope  ($L(E)  =  \frac{d  ln(E\sigma)}{dE}$)
increases  with decrease of energy in contrast to a constant value expected
at the deep sub-barrier energies. Hagino {\it et. al.,} \cite{hagino1} have
pointed out that the failure of the Wong  model  analysis  to  explain  the
unexpected behaviour of fusion cross sections shown by Jiang {\it et. al.,}
\cite{ninisub}  is  partly due to the parabolic shape of the fusion barrier
assumed in the Wong model. Further, Hagino {\it  et.  al.,}  \cite{hagino1}
noted  that  the  increase  in  logarithmic slope with decreasing energy is
consistent with the large diffuseness parameter required to fit the  recent
high  precision  fusion data. In the present work, we have investigated the
unusual behaviour of fusion at  deep  sub-barrier  energies  and  also  the
systematics  of  adiabatic  limit of fusion barriers in terms of the energy
($E$) and angular momentum ($L$)  dependent  barrier  penetration  model  (
{\small{ELDBPM}} ) \cite{eldbpm}.

\par
Owing  to  complexity  of  the {\small{CRC}} calculations, phenomenological
models are needed to gain physical insight of the reaction  mechanism.  The
effective  potentials  derived  from  the  {\small{CRC}}  method provide an
important bridge between the microscopic {\small{CRC}} calculations and the
macroscopic models. It has been shown that the {\small{CRC}}  fusion  cross
sections  and the spin distributions imply an effective fusion barrier that
depends on both $E$ and $L$ through effective radial kinetic energy defined
(see  Eq.  2.)   as   $\epsilon=E-E_{rot}$   \cite{eldbpm}.   Analysis   of
experimental  data  using  $\epsilon$ dependent barrier showed two distinct
limits for the effective fusion potentials. At high energies (above  $E_2$)
well above uncoupled elastic channel barrier, fusion occurs by transmission
through  a  single constant barrier known as sudden fusion barrier ($V_2$).
As  the  energy  decreases  and/or  the  angular  momentum  increases,  the
effective  fusion  barrier decreases linearly with $\epsilon$ until a lower
energy ($E_1$) is reached. At this  lower  limit  called  adiabatic  limit,
fusion  occurs  by  tunneling through a constant barrier known as adiabatic
fusion barrier ($V_1$) \cite{eldbpm,edbpm}. This  simple  effective  fusion
barrier  model was first proposed in \cite{edbpm}, based on the neck degree
of freedom in heavy  ion  fusion.  The  two  dimensional  potential  energy
landscape as a function of relative separation and neck degree exhibits two
distinct  regions  \cite{edbpm}.  At high energies, the colliding nuclei in
the entrance channel fuse by passage over the relative Coulomb  barrier  in
the  radial  space.  As  the  energy  is lowered the fusing system finds an
alternate competing path for fusion by means of neck  formation.  Thus  the
neck formation facilitates the fusion in a dynamic way, which otherwise has
to  fuse  by  passage  through  the  one  dimensional  barrier. At very low
energies, the  effects  of  neck  opening  saturate  and  system  fuses  by
transmission  through  a  lowered  (constant) barrier. The effective fusion
barrier (V$_{eff}$), the transmission factor for fusion ($T_l$), the fusion
cross section ($\sigma_f$), the mean square fusion  spin  ($<L^2>$)  values
and   the   fusion   slope   values  ($L(E)$)  using  the  {\small{ELDBPM}}
\cite{eldbpm} are,
\begin {eqnarray}
V_{eff}(\epsilon) &=& V_2   ~~\mbox{for} ~~~\epsilon\ge E_2  \nonumber\\
V_{eff}(\epsilon) &=&  \alpha \epsilon + \beta   ~~\mbox{for} ~~~E_1\le\epsilon\le E_2  \nonumber\\
V_{eff}(\epsilon) &=& V_1   ~~\mbox{for} ~~~\epsilon\le E_1  \\
\epsilon &=&E-E_{rot}~~~~\mbox{where}~~~E_{rot}=\frac{l(l+1)\hbar^2}{2\mu R_b^2} \\
\sigma_l &=& \frac{\pi}{k^2}(2l+1)T_l(E) \\
T_l(E) &=& \left[1+exp\left(\frac{2\pi}{\hbar\omega}(V_{eff}-\epsilon)\right)\right]^{-1} \\
\sigma_f &=& \sum_l\sigma_l  ~~~\mbox{and}~~~\langle L^2\rangle =\sum_l l(l+1)\sigma_l/\sigma_f \\
L(E)&=&\frac{d\log(\sigma E)}{dE}
\end{eqnarray}
In  Eq.  (1),  $\alpha=(V_2-V_1)/(E_2-E_1)$  and $\beta=-\alpha E_1 + V_1$.
This simple model for fusion consists of all the  physical  ingredients  of
the  {\small{CRC}}  calculations  \cite{eldbpm}.  Further,  the  results of
{\small{EDBPM}} with L-independent barriers \cite{edbpm} can be obtained by
substituting E$_{cm}$ in place of $\epsilon$ dependence  of  V$_{eff}$.  In
the  following,  both  {\small{ELDBPM}} and {\small{EDBPM}} will be used to
understand the heavy ion fusion systematics. In  the  above  equations, the
quantities  $R_b$,  and  $\hbar\omega$  are in general taken independent of
energy. However, these quantities  may  be  different  at  the  sudden  and
adiabatic  limits.  The sudden limit of the barrier is generally around the
one dimensional fusion  barrier,  as  prescribed  by  Vaz  {\it  et.  al.,}
\cite{vaz}  (  see  Eq.  7). The contact radius ($R_{con}$) is the distance
between two touching spheres evaluated by using $r_0$=1.233fm.
\begin {eqnarray}
R_{sud} &=&r_{of}(A_p^{1/3}+A_t^{1/3}) \nonumber\\
r_{ef} &=&2.2951-0.2966\log_{10}(Z_pZ_t) \nonumber\\
r_{of} &=&2.0513-0.2455\log_{10}(Z_pZ_t) \nonumber\\
V_{sud} &=&\frac{Z_pZ_te^2}{r_{ef}(A_p^{1/3}+A_t^{1/3})}
\end{eqnarray}
\par
Wilzynsky  and  Wilzynska  showed  that  the  adiabatic limit of the fusion
barrier can be obtained in terms of the ground state masses of  the  fusing
nuclei  and  the fused system \cite{wilz1}. In this prescription, the depth
of the nucleus-nucleus (NN) potential in the adiabatic limit  is  given  by
the  potential  of  the  compound  nucleus  calculated  with respect to the
potential energy of the two separated incident nuclei of  entrance  channel
with their Coulomb energy ($C_0$ in MeV) subtracted.  Thus, the depth $V_0$
(in  MeV)  of a Woods-Saxon form of the NN potential in the adiabatic limit
is  given  by,  $V_0  =  \left(M_p+M_t-M_{cn}\right)c^2+C_0  $.   Recently,
Wilzynska  and  Wilzynsky  modified  the  above prescription by considering
shell correction term for the  compound  nucleus  \cite{wilz2}.  The  shell
correction  energy  (S$_{cn}$)  is  taken  from the Moller and Nix {\it et.
al.,} \cite{moller}. In general, this term is found to lower the  adiabatic
barriers.  As  mentioned  in  \cite{wilz2},  this  shell correction term is
necessary because it produces only a local dip in the flat landscape of the
nuclear potential  energy  for  the  equilibrium  shapes.  The  diffuseness
parameter  ($a_0$) of the potential can be determined by matching the first
derivative of the NN potential (i.e., NN force) to that  of  the  proximity
potential  at  the  contact distance $R_0$=$R_p+R_t$. A correlation between
the adiabatic barriers and the fusion energy thresholds was shown in  \cite
{wilz2}  over  a  wide  range  of  heavy ion systems. This fusion threshold
energy is defined as the energy at which measured fusion cross  section  is
equal to the s-wave absorption cross section.
\begin     {eqnarray}
V_0  &=& \left(M_p+M_t-M_{cn}\right)c^2+C_0+S_{cn}\\
C_0       &=&      0.7054      \left(\frac{Z_{cn}^2}{A_{cn}^{1/3}}
-\frac{Z_{p}^2}{A_{p}^{1/3}}-\frac{Z_{t}^2}{A_{t}^{1/3}}\right)  \\
V_{NN}(r)&=&\frac{-V_0}{1+exp(\frac{r-R_0}{a_0})}     \\
\left.\frac{dV_{NN}}{dr}\right|_{r=R_0}&=&\frac{-V_0}{a_0}=16\pi\gamma\frac{R_pR_t}{R_p+R_t}
\end{eqnarray}
In  the above equations, $\gamma$ is the surface tension coefficient, $r_0$
is the radius parameter. In  \cite{wilz1,wilz2},  these  are  taken  to  be
$\gamma=0.9516\left[1-1.7826\left(\frac{N-Z}{A}\right)^2\right]  MeV/fm^2$,
$r_0$=1.15fm. The adiabatic barriers  obtained  from  the  prescription  of
\cite{wilz2}  using  these  $\gamma$  and $r_0$ are shown in the Table 1 in
second column (V$_{ad}$) for  various  systems  listed  in  column  1.  The
corresponding barrier radius ($R_b$), barrier curvature ($\hbar\omega$) and
the  potential diffuseness at adiabatic limit are given in Table in columns
3 to 5. The diffuseness parameter of  NN  potential  turns  out  be  around
0.90-1.20  fm  as  shown  in  the  Table.  Such  larger diffuseness values,
reported in some of the fusion studies \cite{mdasgrev}, arise naturally  in
the  prescription  of adiabatic barriers. In order to verify this empirical
prescription of \cite{wilz2}, it is necessary to  determine  the  adiabatic
fusion  barriers  using  experimental  data.  Further, the shell correction
prescription needs to the examined  in  detail  by  studying  the  isotopic
dependence  of adiabatic barriers. Therefore, we have derived the adiabatic
barriers using the {\small{ELDBPM}} fits to the experimental fusion data of
some heavy ion systems with Z$_p$Z$_t$ values ranging from 400 to 1100.\\

\noindent
The        best       fit       parameters       of       {\small{ELDBPM}},
E$_2$,V$_2$,E$_1$,V$_1$,R$_b$,$\hbar\omega$ and the $\chi^2$ values of fits
are listed in columns 10 to 16 of  Table.  The  resulting  fits  to  fusion
excitation  functions  are  shown  in  Figure  1, with solid curves and the
experimental data by symbols for the 26 systems considered  in  this  work.
The    experimental    data    for    these    systems   are   taken   from
\cite{sisni,szr,arsnsm,tizrnb,ninisub,nini5858,nini,nimo,krge}.  The  V$_1$
values  listed in the Table in column 13 are the adiabatic barriers derived
from the experimental data for all these cases. These V$_1$ values  do  not
agree with the adiabatic barriers given in column 2 of Table and therefore,
the  prescription  of  \cite{wilz2}  under predicts the V$_1$ values. It is
observed that  the  values  of  $r_0$  and  $\gamma$  of  \cite{wilz2}  are
different  from  the  values  used in compilations of Moller {\it et. al.,}
\cite{moller},                                                        where
$\gamma=1.25\left[1-2.3\left(\frac{N-Z}{A}\right)^2\right]  MeV/fm^2$ based
on $r_0=1.16 fm$. Therefore, we  have  taken  these  values  of  $r_0$  and
$\gamma$ consistent with nuclear data compilations. The ground state masses
and  shell  correction  energy  are  also  taken  from  \cite{moller}.  The
resulting  empirical  adiabatic   barrier   parameters   V$_{ad}$,   $R_b$,
$\hbar\omega$  and  the  diffuseness  are  listed in the columns 6-9 of the
Table. 1. These V$_{ad}$ values are very  close  to  the  $V_1$  values  of
column  13.  The  isotopic  variation  for  a  given Z$_p$Z$_t$ are grouped
together and are separated by a horizontal line in the Table. The  isotopic
dependence of the adiabatic barriers is very well reproduced in the present
work. The diffuseness parameters are in the range of 0.65-0.9fm as shown in
9$^{th}$  column  of  Table,  in  contrast  to  the  diffuseness  values of
\cite{wilz2} (see column 5). The effective diffuseness of the NN  potential
in  the  adiabatic  limit  is  reduced by our choice of the surface tension
parameter, leading to correct adiabatic barriers \cite{svs1}. Consequently,
the $\hbar\omega$ values in the adiabatic limit are  slightly  larger  than
those of \cite{wilz2} (columns 8 and 4 of Table.)

\par

\noindent
Figure  2(a)  shows  the {\small{ELDBPM}} fits to fusion cross sections for
the $^{58}$Ni+$^{58}$Ni system in solid curves. The  experimental  data  of
\cite{nini5858,ninisub}  is  shown  by circles, showing the good quality of
the model fits. In Figure 2(b) we show the  {\small{ELDBPM}}  fits  to  the
fusion excitation function of $^{60}$Ni+$^{89}$Y. The fusion of this system
has  been  shown  to  exhibit  anomalously steeper fall in deep sub barrier
region as compared to Wong's model estimates and coupled  channels  method.
It  can  be  seen  that  in the {\small{ELDBPM}}, the cross section at deep
sub-barrier region is reproduced. The  best  fit  values  are  respectively
R$_b$=8.75fm  and  $\hbar\omega$=2.75MeV for $^{58}$Ni+$^{58}$Ni system and
R$_b$=8.0fm and $\hbar\omega$=2.50MeV for $^{60}$Ni+$^{89}$Y. These  values
are lower than the values predicted by the prescription of \cite{wilz2} for
these two systems (see Table).\\

\noindent
We  have  calculated  the  exponential  slope parameter given by Eq. 6. The
slope values $L(E)$ for these two systems are  shown  in  Figs.2(c,d).  The
experimental  slope values are well reproduced in the present work as shown
in figure, except at the lowest energies. These maximum slope values  could
not  be  reproduced,  in  spite  of  using  very  low  value  $\hbar\omega$
(=2.5MeV).  Exact  WKB  transmission  in  place  of  transmission   through
parabolic  barrier  would  tend to increase the theoretical model values of
slope at low energy, as shown in \cite{hagino1}.

\noindent
The  percentage  deviations of adiabatic barrier heights ($V_1$) derived in
the present work as compared to the modified prescription  of  \cite{wilz2}
have  been  calculated  for  the  systems  shown  in  Table. The percentage
deviations  of  sudden  barrier  heights  ($V_2$)  are  compared   to   the
phenomenological formulae given by Vaz {\it et. al.,} \cite{vaz}. These are
shown  in  (left side of) Figures 3(a,b) as a function of Z$_p$Z$_t$, which
is a measure of heavyness of the system.  As  shown  in  Figure  3(a),  the
deviations  of  $V_1$  for all these systems are within $2\%$, where as the
deviations of $V_2$ are upto $6\%$. The solid line  drawn  in  the  figures
shows  general trend of Z$_p$Z$_t$ dependence of these deviations, obtained
by a 2$^{nd}$ order polynomial fit.

\par
The optimised $R_b$ values that fit the fusion data of Fig. 1 are listed in
column 14 of the Table. These values have been plotted in Figure 3(c) (left
part)  as a function of Z$_p$Z$_t$. For a comparison, the adiabatic barrier
radii, the sudden values as given by Vaz {\it et. al.,} \cite{vaz} and  the
contact radii ($R_{con}$) are shown by different symbols. For some systems,
the  required  R$_b$ values are much smaller than the sudden, adiabatic and
contact radii values. The solid and dashed curves are  polynomial  fits  to
data  showing  general trends. It should be noted that the {\small{ELDBPM}}
parameters listed in Table are a result of a uniform procedure  applied  to
all systems. Further, the R$_b$ and $\hbar\omega$ values are independent of
energy  and  angular  momentum and therefore they are effective parameters.
These parameters also should be varied between their sudden  and  adiabatic
limit  values  in an $\epsilon$ dependent way, similar to effective barrier
of Eq. (1). Such an implicit E and L dependence of $\hbar\omega$ mimics the
true WKB transmission. \\

\noindent
Similar  analysis  was  performed  using  {\small{EDBPM}}  method where the
effective fusion  barrier  depends  only  on  E$_{cm}$  and  hence  without
L-dependence.  The results of fusion barrier systematics in {\small{EDBPM}}
analysis are shown in right panel of Fig.3. The fits to  fusion  excitation
functions  and  resulting  fusion  slope values of {\small{EDBPM}} are very
similar to the those in Figs. (1,2).

\par
As  shown  in Fig.3(c), the values of R$_b$ derived in the present work are
smaller than the values expected from high energy systematics given by  Vaz
{\it et. al.,}. In order to understand this behaviour, we have analysed the
high  energy  fusion  data  of  a  few systems by a one dimensional BPM. We
performed a three parameter fit to above barrier fusion data by  optimising
V$_b$,R$_b$  and  $\hbar\omega$.  For some systems like $^{36}$S+$^{58}$Ni,
the R$_b$ values turn out to be smaller. For the  $^{58}$Ni+$^{58}$Ni  this
procedure  yields  a slightly higher value than the {\small{ELDBPM}} values
(listed in table). In the high energy region above  E$_2$,  fusion  can  be
obtained  by  $\sigma_f=\pi  R_b^2(1-V_b/E)$.  Therefore,  the smallness of
R$_b$ is a requirement of the high energy fusion cross sections for some of
the systems.\\

\par
\noindent
The  $\hbar\omega$  values  are important for deep sub barrier data and are
closer to the empirical adiabatic values listed in Table. The smallness  of
$\hbar\omega$  is  system  dependent,  especially as shown for the cases of
$^{36}$S+$^{58}$Ni and $^{58}$Ni+$^{58}$Ni, where fusion  data  extends  to
deep  sub  barrier  region.  The  optimised  $\hbar\omega$ values the Ar+Sm
systems are around 5MeV (see Table). Though these systems  have  Z$_p$Z$_t$
values  similar  to Ni+Y system, the barrier curvatures are very different.
These observations may be understood based on neck degree  of  freedom  for
fusion.  In  the  macroscopic  picture,  fusion  occurs by both the barrier
penetration in radial space and neck opening. The barrier curvature depends
on the reduced mass. The mass tensor is a complex quantity and  depends  on
size  of  the system as well as structural details of the colliding nuclei.
The neck formation is more favourable for  heavier  systems  and  the  mass
tensor is smaller, resulting in larger $\hbar\omega$ values. Often, neutron
flow  is  also  considered  to  be  a  classical  analogue of neck opening.
Consequently, the $\hbar\omega$ can be larger  even  for  a  light  system,
whenever neutron flow is favoured.\\

\par
\noindent
In  order to confirm the anomalous small values of R$_b$ and $\hbar\omega$,
study of systematics of fusion at high energy is very important. Therefore,
we used the program {\small{CCFUS}} \cite{ccfus} to  fit  the  high  energy
fusion  data  without any couplings. The parameter dV of {\small{CCFUS}} is
adjusted to fit the cross sections above 100mb for each system (except  for
Kr+Ge systems which could not be reproduced). The resulting barrier height,
R$_b$  and  $\hbar\omega$ and the dV values used are listed in the Table in
the last four columns. It can be seen that  this  high  energy  systematics
from  {\small{CCFUS}}  also  yields smaller R$_b$ and $\hbar\omega$ values.
However, the barrier heights  obtained  by  this  procedure  represent  the
uncoupled  entrance  channel  barriers.  In  the  coupled  channels  method
\cite{cc}, these high energy barriers are also modified by the couplings.\\

\par
\noindent
{\bf Acknowledgements}\\
We  acknowledge  fruitful discussions with B. K. Nayak, R. G. Thomas, P. K.
Sahu and K. Mahata. \\

\par
\noindent
{\bf Conclusion}
\par

\noindent
The  fusion  cross  sections from well above barrier to extreme sub-barrier
energies have been analysed using the energy (E) and angular  momentum  (L)
dependent  barrier  penetration  model  (  {\small{ELDBPM}}  ).  From  this
analysis, the adiabatic limits of fusion barriers have been determined  for
a  wide range of heavy ion systems. The empirical prescription of Wilzynska
and Wilzynski has been used with  modified  radius  parameter  and  surface
tension  coefficient  values  consistent  with  the parameterization of the
nuclear  masses.  The  adiabatic  fusion  barriers  calculated  from   this
prescription are in good agreement with the adiabatic barriers deduced from
{\small{ELDBPM}}  fits to fusion data. The nuclear potential diffuseness is
larger at adiabatic limit, resulting in a lower  $\hbar\omega$  leading  to
increase  of  "logarithmic  slope"  observed  at  energies  well  below the
barrier. The effective fusion  barrier  radius  and  curvature  values  are
anomalously  smaller than the predictions of known empirical prescriptions.
A detailed comparison of the systematics of fusion barrier with and without
L-dependence has been presented.


\begin{center}{
\begin{table}
\begin{tabular}
{|c|c|c|c|c|c|c|c|c|c|c|c|c|c|c|c|c|c|c|c|}
\hline
System &V$_{ad}${\cite{wilz2}}& R$_b$ & $\hbar\omega$ & a$_0$ &  V$_{ad}$ &  R$_b$ &  $\hbar\omega$ & a$_0$ & E$_2$ & V$_2$ & E$_1$ & V$_1$ & R$_b$ & $\hbar\omega$ & $\chi^2$ & V$_{ccf}$ & R$_b$ & $\hbar\omega$ & dV  \\
\hline
$^{28}$Si+$^{58}$Ni   &   50.0 &  10.3 &   3.2 &  0.86 &  52.9 &   9.9 &   3.9 &  0.65 &  56.6 &  55.6 &  51.6 &  52.4 &   9.0 &   3.2 &    1.7  &  55.0  &  9.4  &  3.6 & -5.0   \\
$^{28}$Si+$^{62}$Ni   &   48.6 &  10.6 &   2.5 &  0.90 &  51.6 &  10.2 &   3.7 &  0.68 &  59.0 &  56.8 &  49.5 &  50.5 &   9.2 &   2.5 &    1.9  &  54.1  &  9.7  &  3.7 & 0.0   \\
$^{28}$Si+$^{64}$Ni   &   47.6 &  10.8 &   2.8 &  0.94 &  50.7 &  10.3 &   3.5 &  0.71 &  56.3 &  55.2 &  49.4 &  50.0 &   8.8 &   2.8 &    1.7  &  54.3  &  9.6  &  3.6 & -5.0   \\
\hline
$^{32}$S+$^{58}$Ni    &   56.1 &  10.4 &   3.2 &  0.89 &  59.3 &  10.1 &   3.8 &  0.67 &  62.1 &  61.5 &  58.7 &  59.0 &   8.8 &   3.2 &    5.5  &  64.8  &  9.0  &  2.6 &  -16.0   \\
$^{32}$S+$^{64}$Ni    &   53.1 &  11.0 &   3.5 &  0.99 &  56.5 &  10.6 &   3.4 &  0.75 &  64.5 &  62.5 &  55.7 &  56.5 &   8.8 &   3.5 &   11.3  &  63.8  &  9.1  &  2.5 &  -18.0  \\
$^{36}$S+$^{58}$Ni    &   54.1 &  10.8 &   4.0 &  0.94 &  57.4 &  10.4 &   3.5 &  0.71 &  64.6 &  63.8 &  58.8 &  58.0 &   8.0 &   4.0 &    5.6  &  64.0  &  9.0  &  2.1 &  -19.0  \\
$^{36}$S+$^{64}$Ni    &   52.5 &  11.1 &   2.5 &  0.98 &  55.8 &  10.7 &   3.2 &  0.74 &  63.4 &  59.4 &  54.8 &  56.0 &   8.8 &   2.5 &    4.1  &  61.8  &  9.6  &  3.0 &  -15.0  \\
$^{36}$S+$^{90}$Zr    &   73.2 &  11.4 &   3.2 &  0.98 &  77.2 &  11.1 &   3.5 &  0.75 &  80.4 &  78.8 &  75.8 &  76.6 &  11.2 &   3.2 &   93.0  &  77.6  & 11.2  &  3.9 &  45.0  \\
$^{36}$S+$^{96}$Zr    &   70.7 &  11.8 &   3.0 &  1.05 &  74.8 &  11.4 &   3.2 &  0.80 &  82.0 &  79.8 &  73.3 &  74.1 &  11.8 &   3.0 &  110.8  &  75.6  & 11.5  &  3.8 &  70.0  \\
\hline
$^{40}$Ar+$^{112}$Sn  &   98.2 &  11.9 &   5.0 &  1.06 & 103.2 &  11.6 &   3.5 &  0.81 & 107.4 & 106.4 & 100.7 & 102.5 &   9.5 &   5.0 &   16.9  & 109.9  & 11.0  &  3.7 &  0.0  \\
$^{40}$Ar+$^{116}$Sn  &   97.0 &  12.0 &   5.0 &  1.09 & 102.0 &  11.8 &   3.4 &  0.83 & 115.1 & 113.7 & 100.5 & 102.3 &  11.8 &   5.0 &  117.4  & 109.2  & 11.1  &  3.6 &  0.0  \\
$^{40}$Ar+$^{122}$Sn  &   95.4 &  12.2 &   5.0 &  1.12 & 100.3 &  11.9 &   3.3 &  0.85 & 112.7 & 109.9 &  99.2 & 101.4 &  10.8 &   5.0 &   19.2  & 107.3  & 11.3  &  3.7 &  5.0  \\
$^{40}$Ar+$^{144}$Sm  &  118.3 &  12.2 &   4.0 &  1.08 & 123.6 &  12.0 &   3.5 &  0.83 & 133.1 & 130.1 & 120.1 & 122.3 &   9.5 &   4.0 &   50.5  & 130.3  & 11.5  &  3.8 &  5.0  \\
$^{40}$Ar+$^{148}$Sm  &  116.1 &  12.4 &   5.0 &  1.13 & 121.6 &  12.2 &   3.4 &  0.86 & 131.0 & 129.0 & 116.7 & 120.5 &   9.5 &   5.0 &   35.8  & 129.6  & 11.6  &  3.8 &  5.0  \\
$^{40}$Ar+$^{154}$Sm  &  113.6 &  12.7 &   5.0 &  1.19 & 119.2 &  12.4 &   3.2 &  0.91 & 128.4 & 127.8 & 112.9 & 116.3 &  10.0 &   5.0 &   85.2  & 129.4  & 11.6  &  3.6 &  0.0  \\
\hline
$^{46}$Ti+$^{90}$Zr   &   97.3 &  11.7 &   3.8 &  1.08 & 102.4 &  11.4 &   3.4 &  0.81 & 113.3 & 110.7 &  99.7 & 102.1 &  10.5 &   3.8 &    3.3  & 111.6  & 10.4  &  3.1 &  -10.0  \\
$^{46}$Ti+$^{93}$Nb   &   98.6 &  11.8 &   5.0 &  1.10 & 103.9 &  11.5 &   3.3 &  0.84 & 110.9 & 109.7 & 100.2 & 103.6 &  10.2 &   5.0 &    5.6  & 113.8  & 10.5  &  3.1 &  -10.0  \\
$^{50}$Ti+$^{90}$Zr   &   96.8 &  11.8 &   4.0 &  1.06 & 101.7 &  11.5 &   3.3 &  0.80 & 108.8 & 106.8 & 101.6 & 103.0 &  10.2 &   4.0 &    1.9  & 110.1  & 10.6  &  3.1 &  -10.0  \\
$^{50}$Ti+$^{93}$Nb   &   98.4 &  11.9 &   4.0 &  1.07 & 103.4 &  11.6 &   3.3 &  0.82 & 111.0 & 109.4 & 101.9 & 103.9 &  10.2 &   4.0 &    2.5  & 112.3  & 10.7  &  3.1 &  -10.0  \\
\hline
$^{58}$Ni+$^{58}$Ni   &   91.9 &  11.1 &   2.8 &  0.97 &  96.5 &  10.8 &   3.7 &  0.74 & 104.5 & 103.1 &  94.2 &  95.4 &   8.8 &   2.8 &   13.4  & 103.2  & 10.0  &  3.1 &  -10.0  \\
$^{58}$Ni+$^{64}$Ni   &   87.6 &  11.6 &   4.2 &  1.07 &  92.5 &  11.3 &   3.3 &  0.81 &  99.5 &  98.7 &  91.2 &  93.0 &   8.2 &   4.2 &   13.2  & 101.2  & 10.3  &  3.2 &  -10.0  \\
$^{64}$Ni+$^{64}$Ni   &   86.0 &  11.8 &   3.0 &  1.08 &  90.8 &  11.5 &   3.1 &  0.82 &  98.5 &  97.5 &  90.5 &  91.5 &   9.0 &   3.0 &    2.9  &  99.4  & 10.5  &  3.1 &  -10.0  \\
\hline
$^{60}$Ni+$^{89}$Y    &  118.3 &  11.9 &   2.5 &  1.10 & 124.0 &  11.7 &   3.3 &  0.83 & 135.2 & 133.0 & 123.5 & 125.5 &   8.0 &   2.5 &   43.7  & 135.3  & 10.6  &  2.3 &  -13.0  \\
$^{64}$Ni+$^{100}$Mo  &  121.7 &  12.4 &   4.3 &  1.18 & 127.9 &  12.2 &   3.1 &  0.90 & 142.7 & 142.1 & 128.6 & 130.6 &  11.0 &   4.3 &    2.9  & 140.6  & 11.1  &  2.8 &  -10.0  \\
$^{86}$Kr+$^{70}$Ge   &  119.9 &  12.3 &   3.5 &  1.19 & 126.1 &  12.1 &   3.0 &  0.91 & 141.8 & 139.0 & 123.8 & 126.2 &   8.0 &   3.5 &   81.6  & 140.9  & 10.6  &  1.4 &  -15.0  \\
$^{86}$Kr+$^{76}$Ge   &  119.6 &  12.4 &   4.5 &  1.16 & 125.4 &  12.2 &   3.0 &  0.89 & 132.9 & 131.9 & 124.3 & 126.5 &   8.0 &   4.5 &  133.8  & 138.9  & 10.8  &  1.8 &  -15.0  \\
\hline
\end{tabular}
\caption{The  fusion  barrier parameters in the adiabatic limit for various
heavy ion systems obtained  from  the  prescription  of  \cite{wilz2}.  The
barrier  quantities  in  the  2$^{nd}$-5$^{th}$  columns are for parameters
$r_0$=1.15fm,  ~$\gamma$=0.9516(1-1.7826I$^2$).  The  results  obtained  by
using   $r_0$=1.16fm,   ~$\gamma$=1.25(1-2.3I$^2$)   are   given   in   the
6$^{th}$-9$^{th}$  columns,  with  I=$\frac{N-Z}{A}$.  The   experimentally
derived  barrier  parameters  using {\small{ELDBPM}} analysis, E$_2$,V$_2$,
E$_1$, V$_1$, R$_b$, $\hbar\omega$ and the $\chi^2$ values  are  listed  in
columns  10  to  16.  The  parameters  V$_b$,  R$_b$,  $\hbar\omega$ and dV
obtained by fitting  high  energy  fusion  data  i.e.,  above  100mb  using
{\small{CCFUS}} program are given in last four columns. In the table, R$_b$
and a$_0$ are in units of fm and the others are in MeV. }
\end{table}
}\end{center}
\begin{figure}
\includegraphics[width=14.0cm,height=20.0cm]{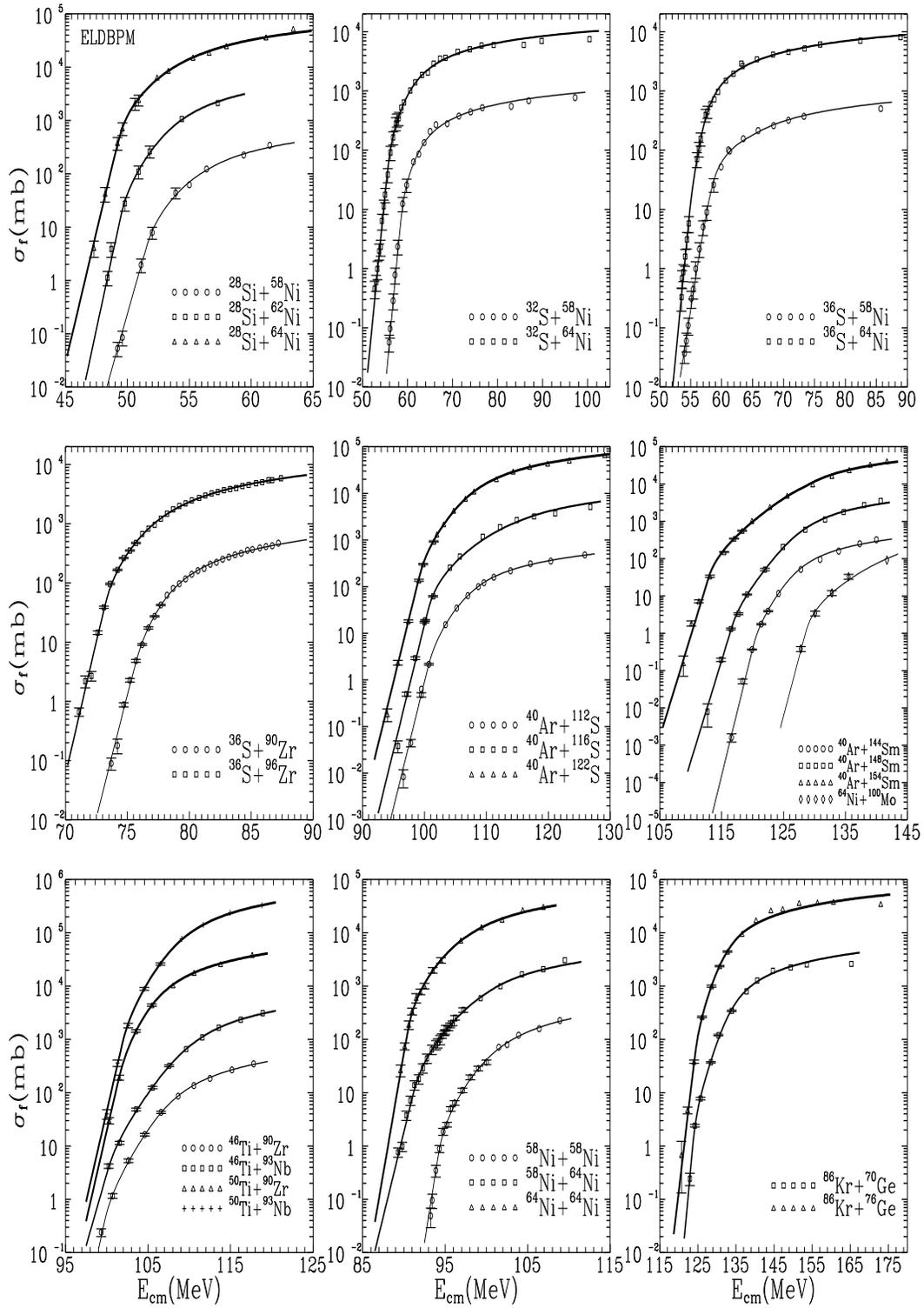}
\caption{Fusion excitation functions for the systems shown in Table. 1. For
clarity, the fusion data of systems within a graph inset  are  successively
scaled  up  by  a decade. The experimental fusion data are shown by symbols
and the solid curves are {\small{ELDBPM}} fits.}
\end{figure}
\begin{figure}
\includegraphics[width=14.0cm,height=12.0cm]{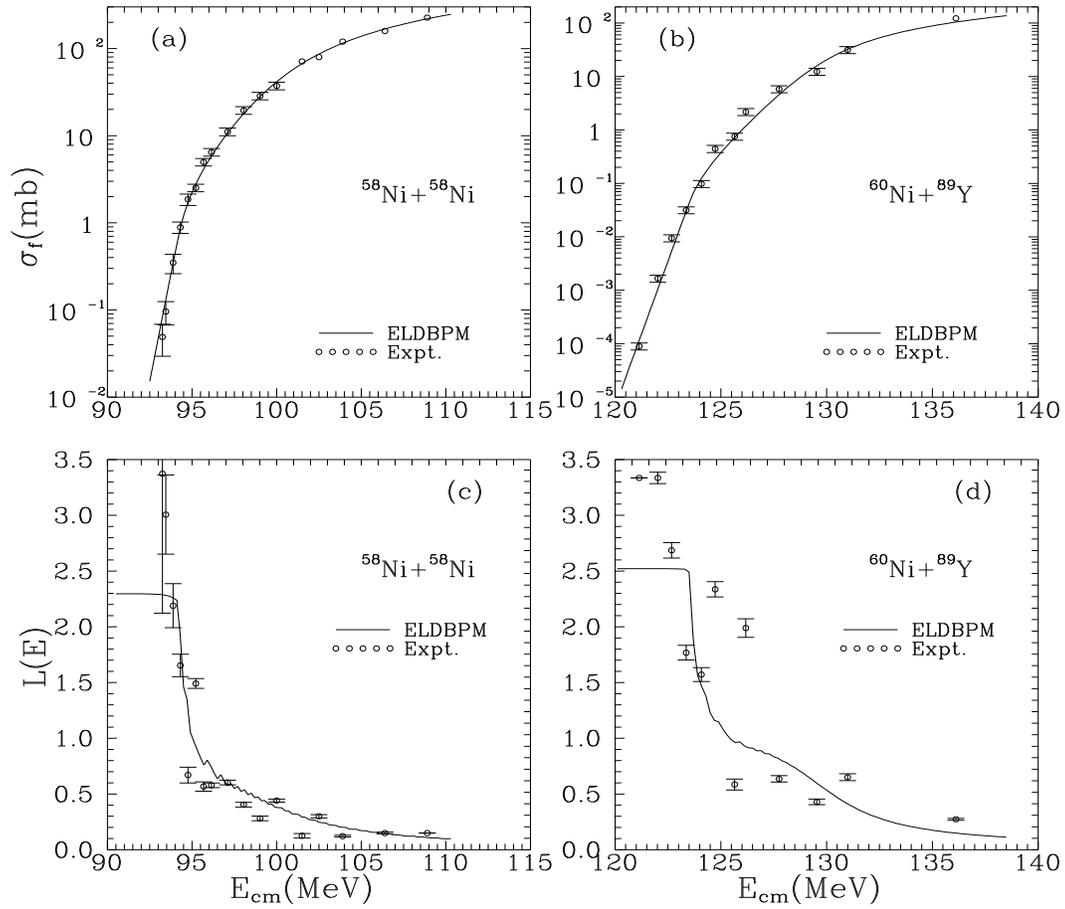}
\caption{Fusion  excitation  function  for  (a) $^{58}$Ni+$^{58}$Ni and (b)
$^{60}$Ni+$^{89}$Y systems. The  experimental  fusion  data  are  shown  by
circles  and  the  solid  curves  are  {\small{ELDBPM}}  fits. Fusion slope
function defined in {\cite{ninisub}} for these two  systems  are  shown  in
Figs.  (c,d) respectively . The experimental fusion slope data are shown by
circles with error bars and the solid curves are {\small{ELDBPM}} results.}
\end{figure}
\begin{figure}
\includegraphics[width=14.0cm,height=20.0cm]{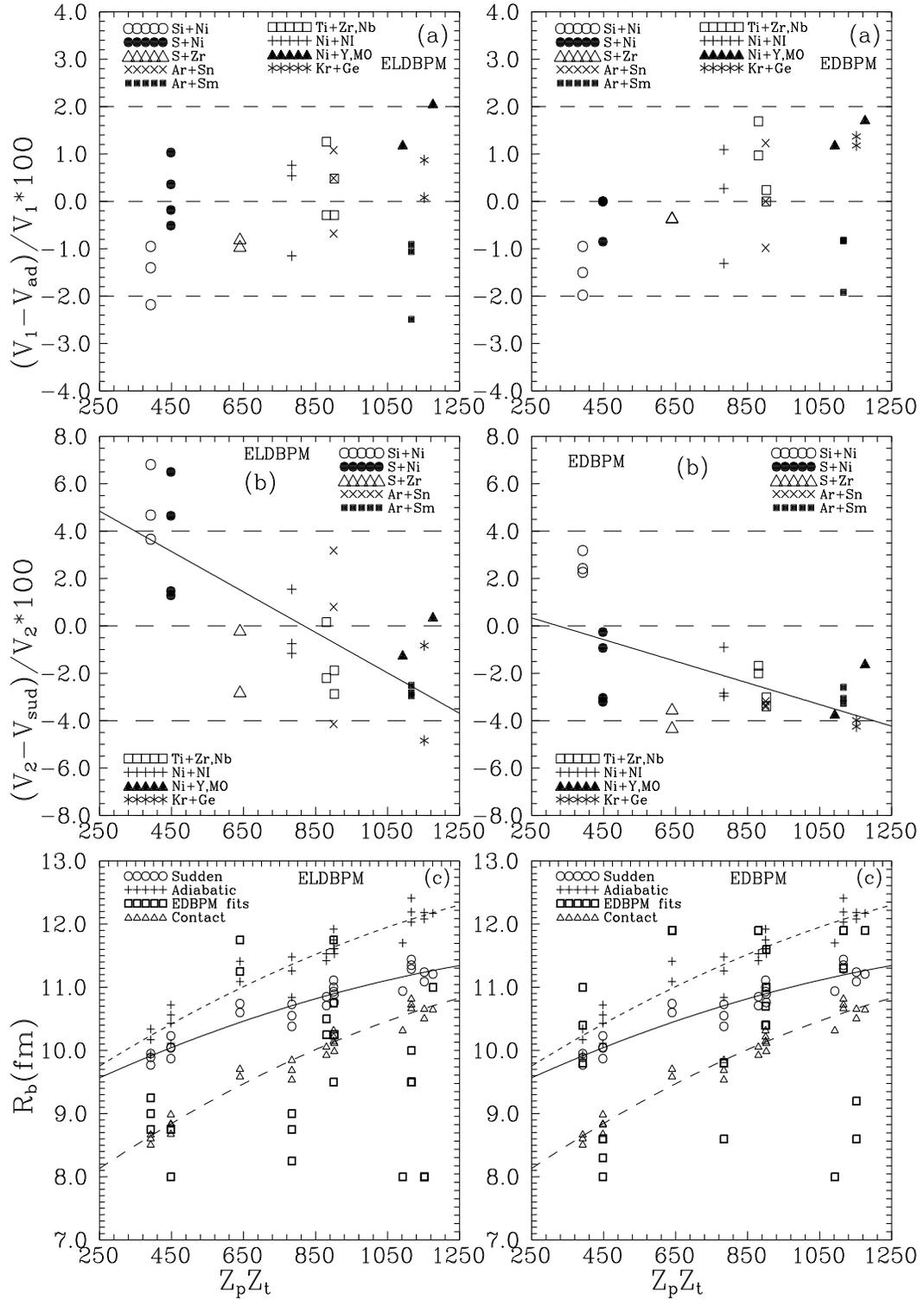}
\caption{ Percentage deviation of the (a) adiabatic and (b) sudden barriers
obtained   from  empirical  prescriptions  and  the  experimental  barriers
obtained using  {\small{ELDBPM}}  method.  The  systems  shown  by  various
symbols  are  for various projectiles listed in the table. In part (c), the
fusion  barrier  radius  values   obtained   from   {\small{ELDBPM}},   the
corresponding  sudden  values from Vaz et al., the adiabatic values and the
contact distance are shown for a comparison.  The solid and dashed lines in
the figures (a)-(c) are guides to eye as mentioned in the text.}
\end{figure}
\end{document}